\RequirePackage{fix-cm}
\documentclass[onecolumn,epjc3]{svjour3}  
\smartqed  
\RequirePackage{graphicx}
\usepackage{multicol}
\usepackage{multirow}
\usepackage{tikz,xcolor,hyperref}
\RequirePackage{float}
\RequirePackage{hyperref}
\RequirePackage{amsmath}
\RequirePackage{amssymb}
\usepackage{widetext}
\RequirePackage{mathtools}  
\usepackage{caption}
\journalname{}
\begin{document}
\definecolor{lime}{HTML}{A6CE39}
\DeclareRobustCommand{\orcidicon}{%
	\begin{tikzpicture}
	\draw[lime, fill=lime] (0,0) 
	circle [radius=0.20] 
	node[white] {{\fontfamily{qag}\selectfont \tiny ID}};
	\draw[white, fill=white] (-0.0625,0.095) 
	circle [radius=0.012];
	\end{tikzpicture}
	\hspace{-2mm}
}

\foreach \x in {A, ..., Z}{%
	\expandafter\xdef\csname orcid\x\endcsname{\noexpand\href{https://orcid.org/\csname orcidauthor\x\endcsname}{\noexpand\orcidicon}}
}
\newcommand{\orcidauthorA}{0000-0002-7350-7043}
\title{BBN Constraints on $f(Q,T)$ Gravity}
%\subtitle{Do you have a subtitle?\\ If so, write it here}

%\titlerunning{Short form of title}        % if too long for running head

\author{Snehasish Bhattacharjee \orcidA{} \thanksref{e1,addr1}                
      }

%\thankstext{t1}{Grants or other notes
%about the article that should go on the front page should be
%placed here. General acknowledgments should be placed at the end of the article.
\thankstext{e1}{e-mail: snehasish.bhattacharjee.666@gmail.com}

%\authorrunning{Short form of author list} % if too long for running head

\institute{Department of Physics, Indian Institute of Technology, Hyderabad 502285, India  \label{addr1}           
     }

\date{\today}

\maketitle

\begin{abstract}
$f(Q,T)$ gravity is a novel extension of the symmetric teleparallel gravity where the Lagrangian $L$ is represented through an arbitrary function of the nonmetricity $Q$ and the trace of the energy-momentum tensor $T$ \cite{fqt}. In this work, we have constrained a widely used $f(Q,T)$ gravity model of the form $f(Q,T) = Q^{n+1} + m T$ from the primordial abundances of the light elements to understand its viability in  Cosmology. We report that the $f(Q,T)$ gravity model can elegantly explain the observed abundances of Helium and Deuterium while the Lithium problem persists. From the constraint on the expansion factor in the range $0.9425 \lesssim Z \lesssim1.1525$, we report strict constraints on the parameters $m$ and $n$ in the range $-1.13 \lesssim n \lesssim -1.08$ and $-5.86 \lesssim m \lesssim12.52$ respectively.

% \subclass{MSC code1 \and MSC code2 \and more}
\end{abstract}

\section{Introduction}

The $\Lambda$CDM cosmological model has been remarkably successful in expounding the dynamics and evolution of the Universe from seconds after the big bang to the present accelerated expansion. However, the model presumes the Universe is statistically homogenous and isotropic at large scales, and that the two mysterious entities termed the dark matter and the dark energy exists in substantial proportions, and that the law of gravity is well described by General Relativity (GR). Nevertheless, the model is incomplete, is evident from the fact that hitherto, no conclusive evidence has been found to confirm the existence of dark matter and dark energy. Furthermore, the model cannot explain the existence of matter-antimatter asymmetry and it is becoming increasingly difficult to incorporate GR into quantum field theory to get a complete theory of reality \cite{1,2,3,4,5,6,7,8,9,10}. Moreover, new models beyond the  $\Lambda$CDM model are being continuously proposed to alleviate numerous tensions between observational data at different scales \cite{11,12,13,14,15,16,17,18}.\\
With this in mind, many alternatives have surfaced which aim to explain the effects of dark matter and dark energy by modifying or extending GR. Extended theories of gravity are motivated by the fact that Einstein's GR does not provide a complete understanding of several gravitational effects at the UV and IR scales. In addition to introducing theories such as the $f(R)$ gravity, which are essentially simple geometric extensions of GR, many alternative theories of gravity identify the gravitational field to be best described by variable(s) other than the Ricci scalar. One of the cornerstones of GR and many extended theories of gravity is the assumption of the Equivalence principle \cite{1} which leads towards coinciding the causal structure and the geodesic and secures the Levi-Civita connection \cite{ift}. \\
An approach to understanding gravity employing variables other than the Ricci scalar could turn out to be useful. For instance, the Ricci scalar could be replaced by the torsion scalar if the gravitational interactions are expressed in terms of tetrads. Such a theory of gravity is commonly termed teleparallel gravity, in which the affinities and not the Equivalence principle plays the most important role. The most successful formulation of a teleparallel gravity is the $f(T)$ gravity which replaces the Ricci scalar $R$ in the Einstein-Hilbert action with the Torsion scalar $T$.  \\
Recently, \cite{fqt} proposed a novel extension of the symmetric teleparallel gravity called the $f(Q,T)$ gravity for which the Lagrangian $L$ is represented through an arbitrary function of the non-metricity $Q$ and the trace of the energy-momentum tensor $T$. The gravitational equations are obtained through the variation of both the metric and the connection with the coupling between matter and geometry accompanying non-conservation of the energy-momentum tensor \cite{fqt}. $f(Q,T)$ gravity has been employed to successfully explain the matter-antimatter asymmetry \cite{b1} and the late-time acceleration \cite{b2}.\\
These alternative theories of gravity are formulated with the primary objective of being self-consistent and furnish a complete picture of the evolution of the Universe from the early phase of inflation to the period of structure formation to the late-time acceleration (readers may refer to\cite{20,21,22} a comprehensive discussion). Cosmography \cite{23,24} and Big-Bang Nucleosynthesis (BBN) \cite{25} are capable of providing powerful constraints on several of these alternate cosmological models. In particular, since the chemical abundances of the primordial light elements such as Deuterium ($D$), Helium ($He$), and Lithium ($Li$) have been ascertained to high precision, BBN offers stringent constraints on alternate theories of gravity since modified theories of gravity must explain the abundances of these metals to validate their applicability and efficiency. \\
In this work, we plan to constrain $f(Q,T)$ gravity from the primordial abundances of the light elements to understand its viability in cosmology. The abundances of the aforementioned light elements have been estimated through various observational techniques. The abundance of Deuterium has been estimated from the absorption lines of gas clouds \cite{34,35,36,37,38}, while that for the Helium from the emission lines of the nearby ionized Hydrogen regions in metal-poor starburst galaxies \cite{30,31} and finally for  Lithium from the atmospheres of very metal-poor stars \cite{32,33}. This technique has been successfully exercised to constrain scalar-tensor gravity \cite{e32}, $f(T)$ gravity \cite{25,ift}, $f(R)$ gravity \cite{e28,e29,e30}, $f(R,T)$ gravity \cite{b3}, Brans-Dicke cosmology with varying $\Lambda$ \cite{e33}, massive gravity \cite{e34} and higher dimensional dilation gravity \cite{e2}. \\
The manuscript is organized as follows: In Sec \ref{sec2} we provide an overview of $f(Q,T)$ gravity. In Sec \ref{sec3}, we delineate the concept of Big Bang Nucleosynthesis and constrain the model parameters of the $f(Q,T)$ gravity model, and in Sec \ref{sec4} we present the conclusions.

\section{Overview of $f(Q,T)$ Gravity}\label{sec2}

The action in $f(Q, T)$ gravity reads \cite{fqt}
\begin{equation} \label{1}
S =\frac{1}{16 \pi} \int \sqrt{-g} \left[   f(Q, T) + \mathcal{L}_{M} \right] d^{4}x
\end{equation} 
where $g$ denote the metric scalar, and $\mathcal{L}_{M}$ denote the matter Lagrangian. \\
Varying \eqref{1} with respect to the metric tensor generates the following field equation \cite{fqt}
\begin{widetext}
\begin{equation}\label{3}
8 \pi T_{\epsilon\varepsilon} =  f_{ T} \left(  T_{\epsilon\varepsilon} + \Theta_{\epsilon\varepsilon}\right) +f_{Q} \left(2 Q^{\alpha \beta }_\epsilon P_{\alpha \beta \varepsilon} -P_{\epsilon \alpha \beta} Q_{\varepsilon}^{\alpha \beta}   \right)  -\frac{2}{\sqrt{-g}}\bigtriangledown_{\alpha} \left( f_{Q}\sqrt{-g} P^{\alpha}_{\epsilon\varepsilon} \right)  - \frac{1}{2} f g_{\epsilon\varepsilon} 
\end{equation}
\end{widetext}
where,
\begin{equation}\label{4}
T_{\epsilon\varepsilon} = -\frac{2}{\sqrt{-g}} \frac{\delta(\sqrt{-g}\mathcal{L}_{M})}{\delta g^{\epsilon\varepsilon}}, \hspace{0.2in}  f_{\epsilon} = \frac{\partial f}{ \partial \epsilon}, \hspace{0.2in} \Theta_{\epsilon\varepsilon} = g^{\epsilon\varepsilon} \frac{\delta  T_{\epsilon\varepsilon}} {\delta g^{\epsilon\varepsilon}}
\end{equation}
and the superpotential $P^{\alpha}_{\epsilon\varepsilon}$ is given as \cite{fqt}
\begin{equation}\label{5}
P^{\alpha}_{\epsilon\varepsilon} = \frac{1}{4}\left[Q ^{\alpha} g_{\epsilon\varepsilon} +2 Q ^{\alpha}_{(\epsilon\varepsilon)} - \delta^{\alpha} _{(\epsilon Q \varepsilon)}- Q^{\alpha}_{\epsilon\varepsilon}   - \tilde{Q}^{\alpha} g_{\epsilon\varepsilon}\right] 
\end{equation}
where \begin{equation}\label{6}
Q_ \alpha = Q _{\alpha \varepsilon} ^{\varepsilon}, \hspace{0.25in} \text{and} \hspace{0.25in}  \tilde{Q}_{\alpha} = Q^{\epsilon}_{\alpha \epsilon}.
\end{equation}
Let us consider a FLRW geometry of the form
\begin{equation}\label{7}
ds^{2} = -N^{2}(t)dt^{2}+ a^{2}(t)\sum_{j=1,2,3} \left(dx^{j} \right) ^{2}
\end{equation} 
where $N(t)$ denote the lapse function, and $a(t)$ denote the scale factor. It may be noted that the lapse function is unity for a FRW background. \\
Substituting \eqref{7} in \eqref{3}, the Friedman equations in $f(Q,T)$ gravity reads \cite{fqt} 
\begin{equation}\label{8}
8 \pi \rho = - \frac{2 \tilde{G}}{1 + \tilde{G}} \left(\dot{F}H + F \dot{H}\right) + \frac{f}{2}  -6 F H^{2}, 
\end{equation}
and
\begin{equation}\label{9}
8 \pi p = 2 \left(\dot{F}H + F \dot{H}\right)- \frac{f}{2} +  6 F H^{2}, 
\end{equation}
where, \begin{equation}
\tilde{G} = \frac{f_{T}}{8 \pi}, \hspace{0.25in} F = f_{Q} .
\end{equation}

For this work, we shall set the functional form of $f(Q,\mathcal{ T})$ to the following \cite{fqt}
\begin{equation}\label{10}
f (Q, T) = Q ^{n+1} + m  T
\end{equation}
where $n$, and $m$ are free parameters. Using \eqref{10}, \eqref{8}, and \eqref{9}, the expression of Hubble parameter $H(t)$ is given as \cite{fqt}
\begin{equation}\label{21}
H (t) = \frac{H_{0} \left(16 \pi - m (\gamma - 4) \right) (n+1) }{3 \gamma H_{0}(t-t_{0}) (m + 8 \pi)  - \left(m \gamma - 4 (\beta + 4 \pi) \right)(n+1) },
\end{equation}
where $H_{0}$, and $t_{0}$ denote respectively the present value of the Hubble parameter and the current age of the Universe, and $\gamma$ is the barotropic EoS parameter.
\section{Big Bang Nucleosynthesis in $f(Q,T)$ Gravity}\label{sec3}

We shall now attempt to constrain the model parameters $m$ and $n$ of the $f(Q,T)$ gravity model from the primordial abundances of the light elements (i.e, $D$, $He$, and $Li$). We are restricting the analysis to a radiation-dominated Universe ($i.e, \gamma = 4/3$) since the phenomena of BBN transpired when the Universe was dominated by radiation. The main idea behind this technique is to obtain a suitable parameter range of $m$, and $n$ for which the theoretical primordial abundances could be consistent with observations. To be more precise, we are concerned with the ratio of the Hubble parameter \ref{21} obtained for the $f(Q,T)$ gravity model ($H_{f(Q,T)}$) to the Hubble parameter for the standard cosmological model (i.e, the Hubble parameter of GR ($H_{GR}$)) in the radiation dominated Universe. We shall define the ratio as 
\begin{equation}\label{z}
Z = \frac{H_{f(Q,T)}}{H_{GR}}.
\end{equation}
It may be noted that the primordial abundances of the light elements are highly sensitive to the baryon density and on the rate of the expansion of the Universe (i.e, on the Hubble parameter) \cite{e37,e38}. The baryon density is expressed as 
\begin{equation}
\eta_{10}=10^{10}\eta_{B}=10^{10}\frac{\eta_{B}}{\eta_{\gamma}},
\end{equation}
where, $\frac{\eta_{B}}{\eta_{\gamma}}$ denote the baryon-to-photon ratio and $\eta_{10}\simeq 6$ \cite{e39}.\\
$Z\neq1$ symbolizes the expansion of the Universe to be governed by a non-standard cosmological model and indicates a non-standard expansion factor. Such cases could transpire if GR turns out to be an incorrect theory of gravity or if there exists additional species of neutrinos other than the three standard types confirmed observationally. However, in this work, we are interested in exploring GR modifications and therefore we shall set the total neutrino species to be equal to three.
 
\subsection{Helium ($^{4}He$) abundance in $f(Q,T)$ gravity} 

The production of Helium is a three step process. In the first step, a Deuterium ($^{2}De$) atom is produced through a neutron ($n$) and a proton ($p$). This is followed by the production of a lighter Helium isotope ($^{3}He$) atom along with a Tritium ($^{3}T$). the equations can be illustrated as follows:
\begin{equation}
n + p \rightarrow ^{2}De + \gamma; \hspace{0.15in} ^{2}De  + ^{2}De \rightarrow {^{3}He} + n; \hspace{0.15in}  ^{2}De + ^{2}De \rightarrow {^{3}T} + p
\end{equation}
Finally, the the heavier helium atom $^{4}He$ is produced through $^{3}T$, $^{2}De$, and $^{3}He$ as follows:
\begin{equation}
^{3}T + ^{2}De \rightarrow ^{4}He + n; \hspace{0.15in} ^{3}He + ^{2}De \rightarrow ^{4}He + p.
\end{equation}
The numerical best fit equation to estimate the primordial abundance of $^{4}He$ can be expressed as \cite{e40,e41}
\begin{equation}
Y_{p} = 0.2485 \pm 0.0006 +0.00016 \left[100(Z-1) + (\eta_{10}-6) \right]. 
\end{equation}
For a standard expansion factor ($i.e, Z=1$), the primordial abundance of $^{4}He$ reads $Y_{p}=0.2485\pm 0.0006$ while astrophysical observations ascertained the same to lie in the range $0.2449 \pm 0.0040$ \cite{e42}. Therefore, by setting $\eta_{10}=6$ we can write,
\begin{equation}
0.2449 \pm 0.0040 = 0.2485 \pm 0.0006 + 0.0016 \left[ 100 (Z-1)\right]. 
\end{equation}
Upon solving this, the constraint on the expansion factor reads $Z = 1.0475 \pm 0.105$.

\subsection{Deuterium ($^{2}H$) abundance in $f(Q,T)$ gravity} 

The production of a Deuterium atom takes place via a neutron and a proton as follows
\begin{equation}
n + p \rightarrow ^{2}D + \gamma.
\end{equation}
The primordial abundance of Deuterium is estimated from the following numerical best fit equation \cite{e37}
\begin{equation}
y_{Dp} = 2.6 \left( 1 \pm 0.0600\right) \left[ \frac{6}{\eta_{10} - 6(Z-1)}\right]^{1.6}.  
\end{equation}
Similar to the previous case, we set $\eta_{10} = 6$. For standard expansion factor $Z=1$, we end up with a theoretical estimate of $y_{Dp} = 2.6 \pm 0.1600$. however, observations reveal the constraints on Deuterium to lie in the range $y_{Dp} = 2.550 \pm 0.0300$ \cite{e42}. Therefore, plugging in the numbers, we end up with the following equation 
\begin{equation}
2.550 \pm 0.030 = 2.60 (1 \pm 0.0600)\left[ \frac{6}{\eta_{10} - 6(Z-1)}\right]^{1.6}.  
\end{equation}
The solution to this equation imposes a strict constraint on $Z$ as $Z = 1.0620 \pm 0.4440$.

\subsection{Lithium ($^{7}Li$) abundance in $f(Q,T)$ gravity} 

The primordial abundance of Lithium is inconsistent with the theoretical predictions of the $\Lambda$CDM model. The abundance of Lithium is estimated to lie between 2.4 to 4.3 times the theoretical predictions \cite{e2,e43}. The baryon density parameter $\eta_{10}$ which describes elegantly the abundances of both Helium and Deuterium, fails to predict the same for Lithium. This is sometimes termed the Lithium problem and hints at the existence of new physics beyond the standard model \cite{e2}.\\
The numerical best-fit equation describing the Lithium abundance can be expressed as \cite{e37}
\begin{equation}
y_{Lip} = 4.82 (1 \pm 0.100) \left[\frac{1}{6}\left\lbrace 3 (1-Z) + \eta_{10} \right\rbrace   \right] ^{2}.
\end{equation} 
The observational abundance of Lithium lie in the range $y_{Lip}= 1.600 \pm 0.300$ \cite{e42}. Plugging this into the above equation constraints $Z$ in the range $Z = 1.960025  \pm 0.076675$.

\subsection{Results}

It is evident that the constraints on the expansion factor for the Helium and Deuterium abundances are similar and would allow the parameters $m$ and $n$ to be consistent with the observations of these elements. However, the $f(Q,T)$ gravity model cannot explain the Lithium abundance given the large discrepancy between observations and theoretical predictions. From Table \ref{table}, we find that the theoretical estimate for the abundances of Helium and Deuterium for the $\Lambda$CDM and the $f(Q,T)$ gravity model are fairly close and falls well within the current observational constraints and therefore allows the parameters $m$ and $n$ to be constrained strictly. \\
We find that for $0.9425 \lesssim Z \lesssim1.1525$, the constraints on the parameters $m$ and $n$ are $-1.13 \lesssim n \lesssim -1.08$ and $-5.86 \lesssim m \lesssim12.52$ respectively.

\captionof{table}{The theoretical predictions for the abundances of $^{2}He$, $^{2}H$ and $^{7}Li$ in $\Lambda$CDM model and in $f(Q,T)$ gravity model along with observational constraints.
}\label{table}
\begingroup
\setlength{\tabcolsep}{10pt} % Default value: 6pt
\renewcommand{\arraystretch}{1.5} % Default value: 1
\begin{tabular}{ |p{4cm}||p{3.0cm}|p{3.0cm}|p{3.0cm}|  }
 \hline

 Models/Observations    & $Y_{p}$ &$y_{Dp}$&$y_{Lip}$\\
 \hline
 \hline
 Observational data &$0.2449 \pm 0.0040$ \cite{e42} & $2.550 \pm 0.030$ \cite{e42} & $1.600\pm 0.300$ \cite{e42}\\
 \hline
  $f(Q,T)$ Gravity&   $0.2487 \pm 0.0174$  & $2.8756 \pm 0.1715$   &$5.29230 \pm 0.5300$\\
 $\Lambda$CDM   & $0.2485 \pm 0.0006$    &$2.600 \pm 0.1600$&   $4.8200 \pm 0.4800$\\

 \hline
 
\end{tabular}
\endgroup
\begin{figure}[H]
  \centering
  \includegraphics[width=8.5cm]{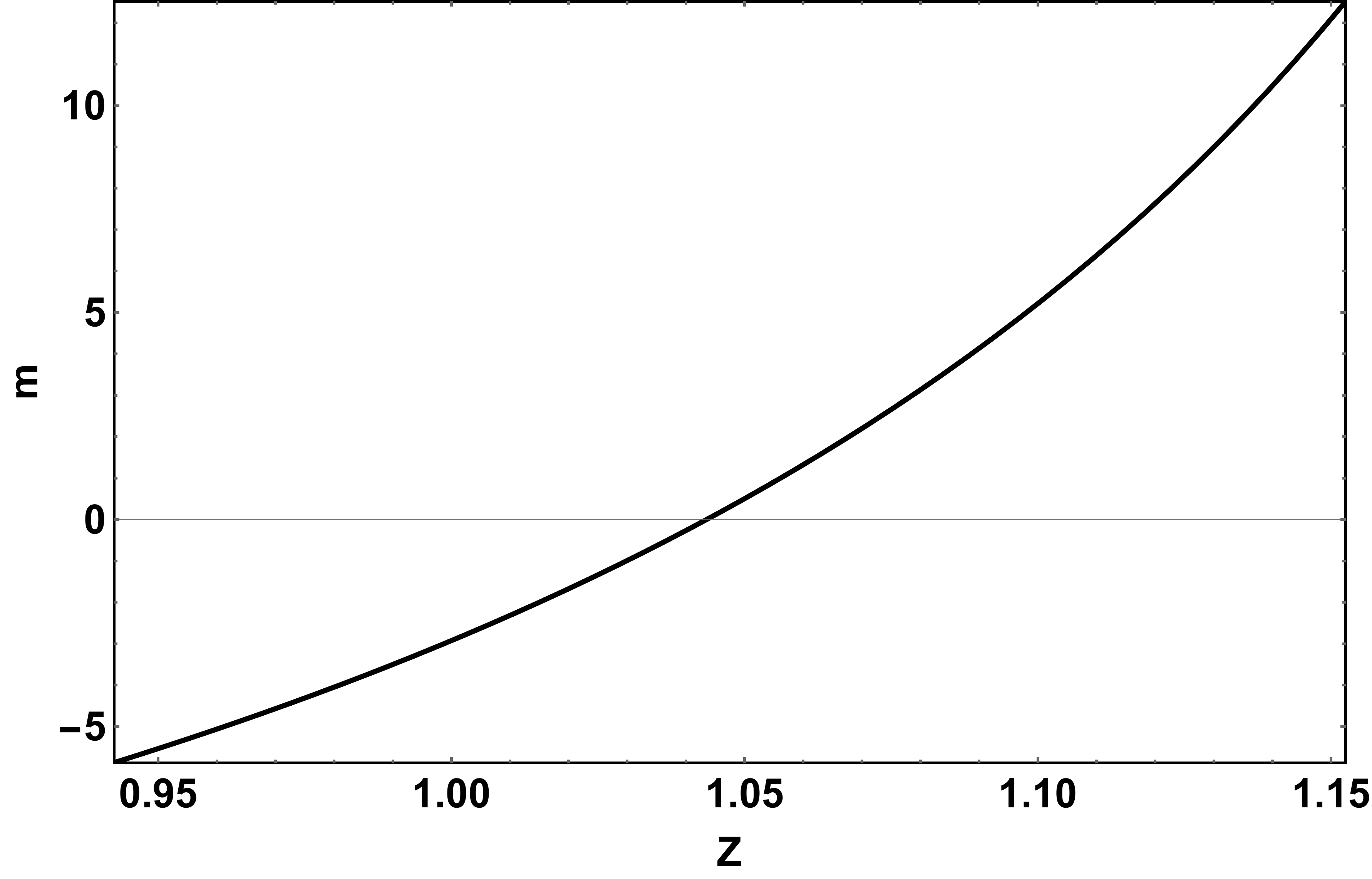}
  \caption{$m$ as a function of $Z$ for a fixed $n=1.11$. The plot is drawn for $0.9425 \lesssim Z \lesssim1.1525$.}
  \label{m}
\end{figure}

\begin{figure}[H]
  \centering
 \includegraphics[width=8.5cm]{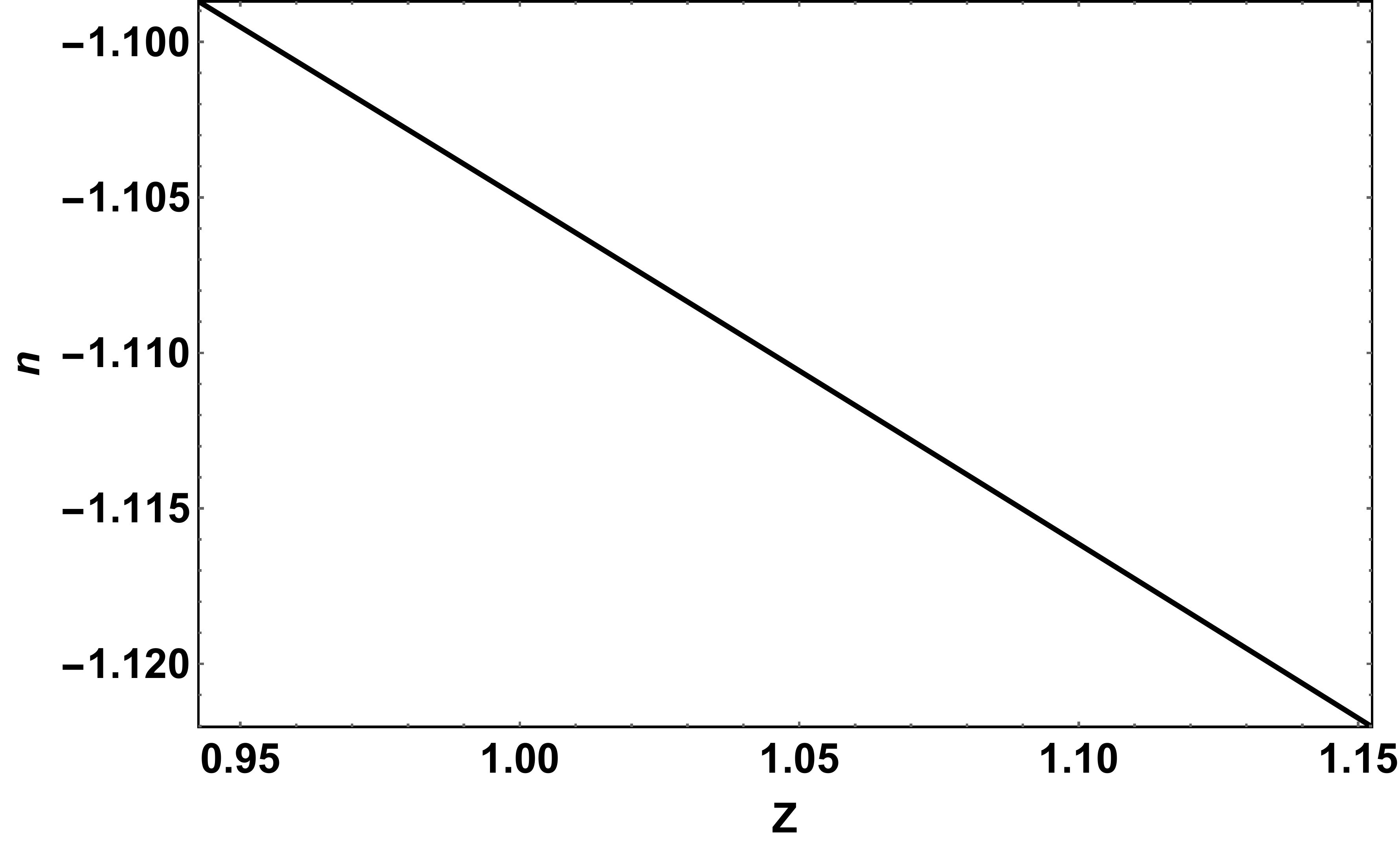}
  \caption{$n$ as a function of $Z$ for a fixed $m=0.1$. The plot is drawn for $0.9425 \lesssim Z \lesssim1.1525$.}
  \label{n}
\end{figure}

\section{Conclusions}\label{sec4}

$f(Q,T)$ gravity is a novel extension of the symmetric teleparallel gravity where the Lagrangian $L$ is represented through an arbitrary function of the nonmetricity $Q$ and the trace of the energy-momentum tensor $T$ \cite{fqt}. $f(Q,T)$ gravity has been very successful in explaining the matter-antimatter asymmetry \cite{b1} and the late-time acceleration \cite{b2}.\\
In this work, we constrained a widely used $f(Q,T)$ gravity model of the form $f(Q,T) = Q^{n+1} + m T$ from the primordial abundances of the light elements to understand its viability in cosmology. We report that the $f(Q,T)$ gravity model explains gracefully the observed abundances of Helium and Deuterium while the Lithium problem persists. From the constraint on the expansion factor in the range $0.9425 \lesssim Z \lesssim1.1525$, we report stringent constraints on the parameters $m$ and $n$ in the range $-1.13 \lesssim n \lesssim -1.08$ and $-5.86 \lesssim m \lesssim12.52$ respectively. Therefore, in addition to explaining the matter-antimatter asymmetry and the late time acceleration, $f(Q,T)$ gravity also explains the abundances of Helium and Deuterium and therefore is turning out to be a viable alternative to the $\Lambda$CDM cosmological model.\\
In future work, we shall try to investigate the growth of density fluctuations and configurational entropy in $f(Q,T)$ gravity to understand the efficiency and applicability in explaining the growth of cosmic structures and the accelerated expansion of the Universe in greater detail.

\section*{Acknowledgments}

I thank Shantanu Desai for reading the manuscript and for the fruitful discussions.
I also thank DST, New-Delhi, Government of India for the provisional INSPIRE
fellowship selection [Code: DST/INSPIRE/03/2019/003141].

\end{document}